\documentclass[conference]{IEEEtran}
\usepackage{latexsym}
\usepackage{graphicx}
\graphicspath{ {./Figures/} }
\usepackage{xcolor}
\usepackage{dsfont}
\usepackage[noadjust]{cite}
\usepackage{amsfonts,amssymb,amsmath}
\usepackage[ruled,vlined]{algorithm2e}
\usepackage[noend]{algpseudocode}
\def\BibTeX{{\rm B\kern-.05em{\sc i\kern-.025em b}\kern-.08em T\kern-.1667em\lower.7ex\hbox{E}\kern-.125emX}}
\DeclareMathOperator*{\argmin}{arg\,min}

\begin{document}

\title{DeepCQ+: Robust and Scalable Routing with Multi-Agent Deep Reinforcement Learning for Highly Dynamic Networks}
\author{\begin{tabular}{cc}
\begin{tabular}[t]{c}
\author{\IEEEauthorblockN{Saeed Kaviani\IEEEauthorrefmark{1}, Bo Ryu\IEEEauthorrefmark{1}, Ejaz Ahmed\IEEEauthorrefmark{1}, Kevin Larson\IEEEauthorrefmark{2}, Anh Le\IEEEauthorrefmark{1}, Alex Yahja\IEEEauthorrefmark{1}, and Jae H. Kim\IEEEauthorrefmark{2}
     \IEEEauthorblockA{\IEEEauthorrefmark{1}EpiSys Science, Inc. : 
     \{saeed, bo.ryu, ejaz, anhle, alex\}@episci.com}\IEEEauthorblockA{\IEEEauthorrefmark{2}Boeing Research and Technology : \{kevin.a.larson, jae.h.kim\}@boeing.com}}}
\end{tabular}
\end{tabular}
}


\maketitle
\thispagestyle{plain}
\pagestyle{plain}
\begin{abstract}
Highly dynamic mobile ad-hoc networks (MANETs) remain as one of the most challenging environments to develop and deploy robust, efficient, and scalable routing protocols. In this paper, we present \textbf{DeepCQ+} routing protocol which, in a novel manner, integrates emerging multi-agent deep reinforcement learning (MADRL) techniques into existing Q-learning-based routing protocols and their variants, and achieves persistently higher performance across a wide range of topology and mobility configurations. While keeping the overall protocol structure of the Q-learning-based routing protocols, \textbf{DeepCQ+} replaces statically configured parameterized thresholds and hand-written rules  with carefully designed MADRL agents such that no configuration of such parameters is required a priori. Extensive simulation shows that \textbf{DeepCQ+} yields significantly increased end-to-end throughput with lower overhead and no apparent degradation of end-to-end delays (hop counts) compared to its Q-learning-based counterparts. Qualitatively, and perhaps more significantly, \textbf{DeepCQ+} maintains remarkably similar performance gains under many scenarios that it was \emph{not} trained for in terms of network sizes, mobility conditions, and traffic dynamics. To the best of our knowledge, this is the first successful application of the MADRL framework for the MANET routing problem that demonstrates a high degree of scalability and robustness even under the environments that are \emph{outside} the trained range of scenarios. This implies that our MARL-based \textbf{DeepCQ+} design solution significantly improves the performance of Q-learning-based CQ+ baseline approach for comparison and increases its practicality and explainability because the real-world MANET environment will likely vary outside the trained range of MANET scenarios. Additional techniques to further increase the gains in performance and scalability are discussed. 
\end{abstract}

\section{Introduction}
Design of a robust and efficient routing algorithm has been one of the most challenging problems in communication and computer networks. This challenge is compounded in tactical MANETs where the network is highly dynamic and unpredictable with respect to mobility and topology,  interference, and possible jamming \cite{elmasry2010comparative}. These factors significantly reduce the reliability, and therefore, many traditional MANET routing protocols require frequent re-computations of end-to-end routes upon detecting network topology changes, resulting in a periodic loss in throughput due to traffic not being sent during the re-computations. This is exacerbated as the rate of topology change increases. To improve packet delivery performance via rapid and efficient exploration in these highly dynamic networks, broadcasting (i.e. transmission of a packet to all neighbors, also known as flooding) has become a popular technique \cite{danilov12,clausen2003optimized, moy1998ospf}. Traditional MANET routing protocols (e.g. OSPF and OLSR \cite{moy1998ospf,clausen2003optimized}) perform reliably when the network is in a stable state but less effectively in highly dynamic networks. Other traditional adaptive strategies also are not responsive fast enough for dynamic networks \cite{danilov12}.


 In this paper, we consider a class of distributed routing algorithms that only share limited information (i.e. two single floating value) through per-hop acknowledgment (ACK) packets. This method of cooperation is efficient as ACKs are inherently present in the networking protocols (such as MAC-layer acknowledgement) and do not require any extra implementation in the system.  The seminal work \cite{Boyan1994} proposed Q-routing, which used a reinforcement learning (RL) module (i.e. Q-Learning \cite{sutton2018reinforcement}) to route packets and minimize delivery time. Each node uses $Q$-values representing quality of paths which are acquired locally to determine the next hop and shared via ACK messages. Kumar et al. \cite{kumar98} improved Q-routing for dynamic networks with the addition of confidence values (i.e. $C$-values) in their \textit{CQ} routing protocol. To improve reliability and exploration speed of the CQ-routing, smart robust routing (SRR) algorithm \cite{johnston18} was proposed to add selective broadcasting actions. SRR utilizes heuristic rules on when to broadcast in order to improve robustness but keep the overall overhead under control. We refer to this technique as \textit{CQ+} routing (also known as Robust Routing for Dynamic Networks, or R2DN) as it is an extension of CQ routing for balancing between reliability and overhead. 
 Although CQ+ routing uses a simple but efficient switching policy to choose between unicast and broadcast, its decisions depend on a single network parameter: best-path confidence level. Consequently, it has a limited snapshot of the entire network that is likely to lead to a locally optimal solution, since it does not fully account for the high rate changes in topology and degree of congestion in likely forwarding paths. Nevertheless, its performance has been consistently producing high delivery ratios across many scenarios used in our study, though at a noticeably high cost of broadcast overhead, and as a result, it serves as a baseline design for our MADRL framework. The question we posed, then, is whether an emerging deep learning framework can help reduce the overhead while maintaining, or exceeding, the performance of CQ-routing (especially goodput).
 
Routing decisions such as a next-hop selection are opportune targets for employing RL-based techniques, as it was originally introduced and initiated by Boyan's Q-routing protocol \cite{Boyan1994}. Following the Q-routing approach, a flurry of new techniques and algorithms from the RL community have been developed and applied to packet routing and scheduling \cite{ali2020hierarchical, yu2018drom, stampa2017deep, ye2019deep, you2020toward}. These techniques are often scale poorly when network sizes increase and system parameters change. To address these shortcomings, new MADRL-based approaches such as \cite{you2020toward} and \cite{ali2020hierarchical} have been proposed, but they suffer from limited scalability due to node-specific policy generated from the training, requiring a re-training every time a new node needs to be introduced to the network. 

To the best of our knowledge, no study has been reported on successfully achieving both scalability and robust performance simultaneously using MADRL in dynamic networks such as MANET. Our \textbf{DeepCQ+} is both scalable and robust by allowing MADRL to control the next-hop adaptive flooding decisions while maintaining the CQ+ routing protocol "structure". In this paper, we describe how a single MADRL-based routing agent is designed and trained to produce a robust and scalable routing policy for any node in the network regardless of the network size. More significantly, our \textbf{DeepCQ+} design and learning methodology enables network designers to train on a limited range of environment parameters (e.g. small network size, selective traffic flow patterns, and limited variations), and still to be effective even when deployed in scenarios outside the trained range of network size, traffic profiles, and mobility patterns. This is an extremely important and unique benefit of \textbf{DeepCQ+} since it does not suffer from the curse of dimensionality due to potentially large network sizes, and catastrophic forgetting even when trained with a wide range of scenarios. 

\section{Robust Routing Framework}

\subsection{SRR (CQ+ routing) Protocol}\label{section:r2dn_protocol}
The SRR algorithm \cite{johnston18} uses the network parameters $C$ and $H$\footnote{We have renamed the original $Q$-factor in CQ-routing to $H$-factor to prevent confusion with the $Q$ in the $Q$-networks and/or $Q$-learning in the RL context.} for the routing decisions primarily introduced in seminal CQ-routing \cite{kumar98}. Each node $i$ has a $H$-factor, $h(i,j,d)$ (i.e. $i \rightsquigarrow j \rightsquigarrow d$), which represents an estimate of the least number of hops between node $i$ and destination $d$ which passes through potential next-hop $j$. To monitor the dynamics of the network, each node $i$ also have a confidence level or $C$-value, $c(i,j,d)$, that represents the confidence in likelihood the packet will reach its destination $h(i,j,d)$. This $C$-value is increased with each packet transmission success (receiving the ACK). Every packet transmission, the $C$-value is degraded by a decay factor. $C$- and $H$-factors are updated through the $c_\text{ack}$ and $h_\text{ack}$ which are propagated by the ACK packets from the receiving node (e.g. next-hop) to the transmitting node. These ACK values are computed at the next-hop node $j$ as 
\begin{equation}\label{h_ack}
h_{\text{ack}} = 1 + h(j,\hat{k},d)
\end{equation}
\begin{equation}\label{c_ack}
c_{\text{ack}} = c(j,\hat{k},d)
\end{equation}
where $\hat{k}$ is the best next-hop estimate of node $j$ to destination $d$ and it is found by $\hat{k} = \argmin\limits_k h(j,d,k)\left(1-c(j,k,d)\right)$
In other words, $C$- and $H$-values are exponential moving average of the $c_\text{ack}$ and $h_\text{ack}$, respectively. If a transmission fails or there is no ACK to update $C$- and $H$-levels, then $H$-level cannot be updated. However, we degrade the $C$-level as in $c_\text{ack} = 0$ to reflect the path failure. The updates of the $C$- and $H$-levels given by 
\begin{equation}\label{update_H}
    h_{t+1}(i,j,d) = (1-\alpha)h_t(i,j,d) + \alpha h_\text{ack},
\end{equation}
\begin{equation}\label{update_C}
    c_{t+1}(i,j,d) =
\begin{cases}
 (1-\lambda)c_t(i,j,d) & \text{failure}\\
 (1-\lambda)c_t(i,j,d) + \lambda c_\text{ack} & \text{otherwise} 
\end{cases}
\end{equation}
where $0 \leq \alpha \leq 1$ is a discount factor for the new observation with an adaptable value of $\alpha = \max\left(c_\text{ack}, 1- c_t(i,j,d)\right)$. $\lambda$ is the decay factor for the new observation of 
$c_\text{ack}$ (or $1-\lambda$ is the decay factor for the old observation). If a packet is received at the destination $d$, then $c_\text{ack}$, and $h_\text{ack}$ are set to 1, indicating that we have full confidence that we are 1 hop away from destination. CQ+ routing algorithm consists of the following three steps:  \textit{(i)} reception of ACK and consequently updating $C$- and $H$-levels; \textit{(ii)} reception of data packets, detection and dropping of duplicate packets, and queuing if not the destination; and \textit{(iii)} transmission of data packets either via unicast or broadcast according to the routing policy chosen. Additional detail of the SRR algorithm (or CQ+ routing) is provided in \cite{johnston18}.  

The CQ+ routing policy chooses the next-hop based on the minimization of the information uncertainty and the expected number of hops. This is given by
\begin{equation}\label{best_next_hop}
    j^\star = \argmin_j h(i,j,d)\left(1-c(i,j,d)\right).
\end{equation}
The next-hop $j^\star$ is considered by the CQ+ routing policy if it unicast the packet to a single neighboring node. However, the CQ+ routing policy triggers broadcast in order to mitigate the next-hop information uncertainty. The CQ+ routing policy is probabilistic and it assigns the probability of broadcast as 
\begin{equation}\label{r2dn_policy}
    P_\text{BC} = \epsilon + (1-c(i, j^\star,d))(1-\epsilon) 
\end{equation}
where a small value $\epsilon$ is used for the minimum probability of broadcast (defined for exploration purposes). In what follows, we briefly describe the MADRL framework employed for replacing the broadcast/unicast decision of the CQ+ routing protocol with a deep neural network. 

\section{Multi-Agent Deep Reinforcement Learning for CQ routing (DeepCQ+)}

\subsection{Background}
Our robust routing design follows the \textit{decentralized partially-observable Markov decision process (Dec-POMDP)}, a popular framework for decision-making problems in a team of cooperative agents \cite{oliehoek2016concise}. Dec-POMDP tries to model and optimize the behavior of agents while considering the uncertainties of both the environment and other agents.

\subsubsection{Policy Optimizations}
Policy gradient (PG) method is another popular choice that is based on the optimization of the policy directly rather than estimating the expected return and it optimizes the policy parameters by descending toward the gradient direction of the expected discounted reward return \cite{sutton2000policy}. Particularly, we deploy a state-of-the-art policy gradient algorithm called Proximal Policy Optimization (PPO) \cite{schulman2017proximal}, which uses multiple epochs of stochastic gradient ascent to perform each policy updates with ease of implementation.

\subsubsection{Centralized Training, Decentralized Execution}
In communication networks, partial observability and communication constraints necessitate the learning of decentralized policies which rely only on local information at each agent. Decentralized policies also avoid the exponentially growing joint action space with the number of agents, and therefore, make them more practical and faster to converge in training. Fortunately, decentralized policies can be learned in a centralized fashion specially in a simulated or controlled environment. The paradigm of \textit{centralized training with decentralised execution} has already attracted a strong attention in the RL community \cite{oliehoek2016concise}. 

\subsubsection{Parameter Sharing}
\begin{figure}[t]
\centering
\includegraphics[width=8.5cm]{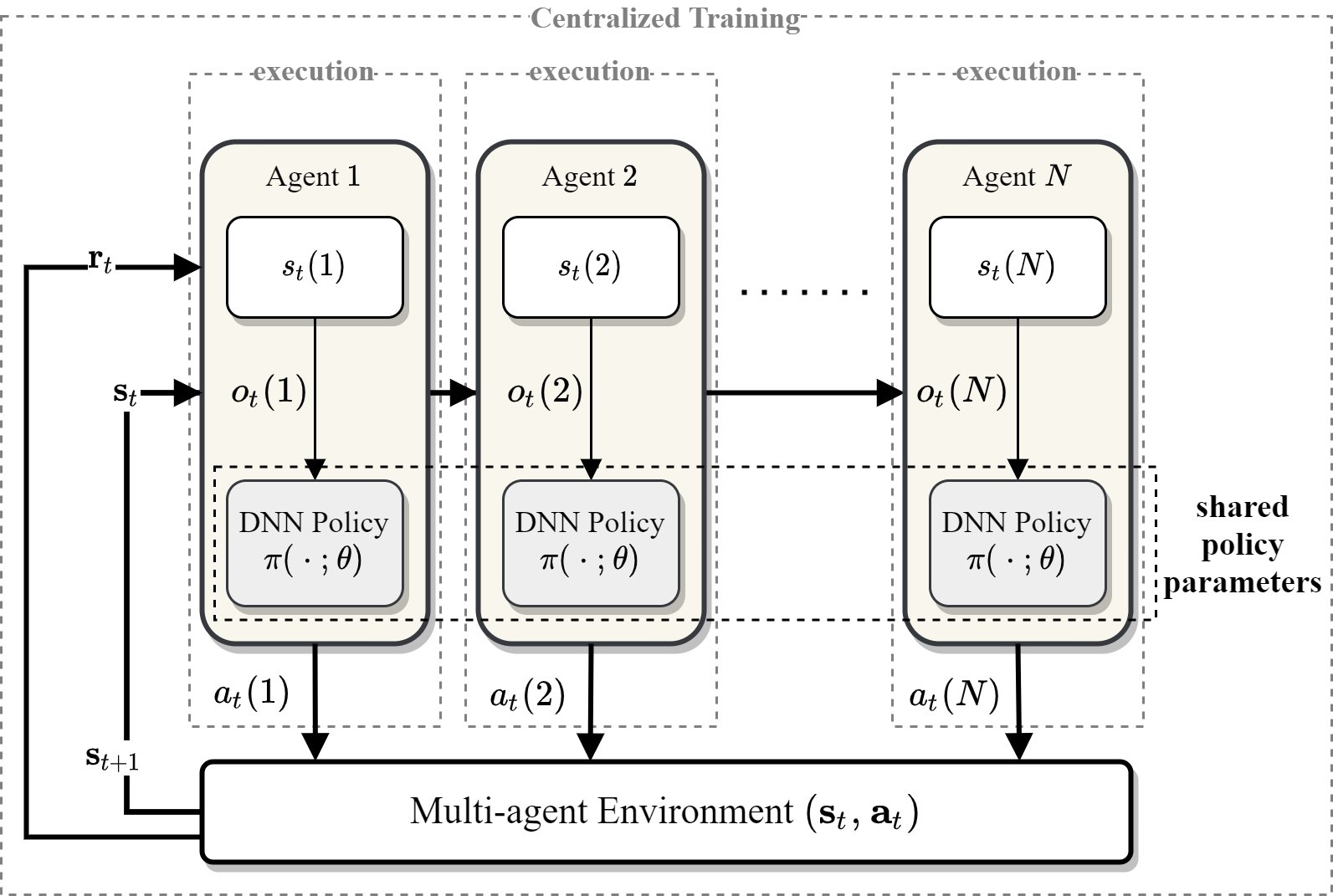}
\caption{Multi-agent network routing environment with shared policy parameters between agents. The centralized training and decentralized execution are also shown. Each agent $i$, uses the shared policy $\pi_\theta$ individually to find its own action $a_i(t)$ based on its own observations $o_i(t)$. The multi-agent environment operates based on the joint-actions decided and taken individually and transition to next state $\mathbf{s}_{t+1}$ and the rewards are pulled out based on that.}
\label{decentralized_exec_centralized_train}
\end{figure}

A common strategy is to share the policy parameters among agents that are homogeneous \cite{terry2020parameter} (parameter sharing). The multi-agent environment for the routing problem and our proposed framework are summarized in Fig. \ref{decentralized_exec_centralized_train} where the centralized training, decentralized execution, and parameter sharing across agents are illustrated.

\subsection{DeepCQ+ Design Framework and Approach}
We consider a homogeneous MANET with variable network size (e.g. $5 \leq N \leq 50$) with multiple unicast traffic flows with randomized source and destination pairs. The nodes are consistently moving at various random velocities and directions. We employ popular MANET mobility models such as Gauss-Markov model and random way-point. As these models are not realistic in general, we enhance these models to provide much more diverse sets of mobility as discussed in Section \ref{sec:env_modelling}. 
Each realization of a network scenario (also called an \textit{episode} in the context of RL) has at least $T$ time-slots (each time step is single packet duration time). For simplicity, it is assumed that packet duration is fixed in time (i.e., a slotted system) but  data rates can still vary. An episode ends once it exceeds a maximum traffic length, $T_{max}$ and all nodes complete their transmissions until queues become empty. No new traffic enters the network after time $T_{max}$. The goodput is computed as the ratio of the total number of successfully delivered packets to the total packets that entered the network. Duplicate packets at the destination are not included. We train and use the same parameterized policy $\pi_\theta$ for every node (one policy for all), which is found to be key to achieve scalability and practicality. When the policy function is defined as a Deep Neural Network (DNN), the parameter vector $\theta$ represents the weights of the neural network.   
We deliberately choose to train our DNN policy with a single network size, single data flow (single source and destination pair), and small range of dynamic levels, and test for varying network sizes with multiple data flows and a wide range of mobility. 

\subsubsection{Pre-processing of the policy input features}
We select only the best $K$ (e.g. $K$ = 4) neighbors out of total $N-1$ possible neighbors in the network for each node according to their available $C$ and $H$ values. For the sake of simplicity, we drop $i$ and destination $d$ from $C$- and $H$-levels and use the next-hop as index (i.e. $c_t(i,j) := c_t(i,j,d) $). Now, in our problem formulation the best $K$ neighbor $C$ and $H$ values are represented by 
\begin{equation}
    \begin{split}
\mathbf{c}_t(i) &= \left[c_t(i,i_1), c_t(i,i_2), \ldots, c_t(i,i_K)\right] \\
\mathbf{h}_t(i) &= \left[h_t(i,i_1), h_t(i,i_2), \ldots, h_t(i,i_K)\right]
    \end{split}
\end{equation}
where the neighboring nodes $i_1,...,i_K$ of $i$ are ordered so that
\begin{equation}
\label{preprocess_CH_sorting}
    h_t(i,i_1)(1-c_t(i,i_1)) \leq \cdots \leq h_t(i,i_K)(1-c_t(i,i_K)).
\end{equation}
\subsubsection{State/Observations}
 We use a fully connected neural network (FCNN)  to capture temporal behavior of network parameters, and  add the change rate between current observations and previous observations as input to the DNN policy. The input features to the DNN policy at node $i$ are given by
\begin{equation}\label{input_features}
    \mathbf{o}_t(i) = \left[\mathbf{c}_t(i), \mathbf{h}_t(i), \Delta\mathbf{c}_t(i), \Delta\mathbf{h}_t(i), a_{t-1}(i), p_{t-1}(i)\right],
\end{equation}
where $\Delta\mathbf{c}_t(i) = \mathbf{c}_t(i) - \mathbf{c}_{t-1}(i)$, $\Delta\mathbf{h}_t(i) = \mathbf{h}_t(i) - \mathbf{h}_{t-1}(i)$, and $a_{t-1}(j)$ is the previous action of some node $j$ from which the current packet is received. The last indicator shows if the current packet is received as a result of another node's broadcast or unicast. 

\begin{table}
\begin{center}
 \begin{tabular}{l  c  c  c } 
 \hline 
 Routing Protocol & C/Q-values & Broadcast & MADRL \\ 
 \hline \hline
 CQ-routing \cite{kumar98} & $\checkmark$ & $\times$ & $\times$ \\ 
 \hline
SRR (CQ+) \cite{johnston18} & $\checkmark$ & $\checkmark$ & $\times$ \\
 \hline
  DeepCQ+ routing (this work) & $\checkmark$ & $\checkmark$ & $\checkmark$   \\
  \hline 
\end{tabular}
\end{center}
\caption{Comparison of robust routing protocols in dynamic networks}
\end{table}
\subsection{Reward Definition for DeepCQ+ routing}
We consider a stochastic routing policy $\pi(a|s_t)$ that sets the rewards for unicast and broadcast as the probability of those actions as follows:
        \begin{equation}
        \label{SRR_rewards}
            r_t(\mathbf{s}_t, a_t) = 
            \begin{cases}
             1-c_t(i,i_1)\tilde{\epsilon} & a_t = 1(\text{broadcast}) \\
             c_t(i,i_1)\tilde{\epsilon} & a_t = 0(\text{unicast})
            \end{cases}
        \end{equation}
where $\tilde{\epsilon} = 1- \epsilon$. We expect that the performance of this RL problem be close to the CQ+ routing (at least for small values of $\gamma$) as both DeepCQ+ and CQ+ have similar rewards. We refer to the above reward as a DNN-based version of the CQ+ routing. Note that different reward designs will produce different performance objectives, thus providing a higher degree of design flexibility compared to the baseline CQ+ routing. 

Note that the above reward specification does not necessarily make the CQ+ routing policy learn to minimize the network overhead. To address this, We first define the overhead as the ratio of the total number of transmissions in the communication network, denoted by $N_\text{TX}$, to the total number of packets delivered, denoted by $N_D$ for a specific packet data rate and a window of time. With this, we define the \textit{normalized overhead} by the network size, $N$, as $\mathsf{OH} = \frac{1}{N} \cdot \frac{N_\text{TX}}{N_\text{D}}$. 
Based on this, we introduce two types of overhead: (i) overhead defined as excess transmitted bits per delivered bits (overhead-1); and (ii) and total transmitted bits (both data and ACK) divided by the total delivered data bits (overhead-2). In order to ensure that the learned policy attempts to minimize the overhead, we define a new reward function as follows:
\begin{equation}
r_t = w_1\mathds{1}_{\mathcal{D}}-w_2\mathds{1}_{\mathcal{Z}} -w_3\frac{N_\text{ack}}{N}
\end{equation}
where $\mathds{1}_{\mathcal{D}}$ is the reward for packet delivery. If a packet is delivered successfully to the destination and node $i$ has contributed to that delivery then it will be rewarded. Also, $\mathds{1}_{\mathcal{Z}}$ is a reward (penalty) indicator which is enabled when we have not received any ACKs from our transmissions. The next term in the reward, i.e. $N_\text{ack}/N$, is the normalized number of ACKs received as a result of the action taken and therefore closely represents the number of copies of a packet at other nodes and the system overhead. The weights of the reward components, (i.e. $w_1, w_2, w_3$), have been tuned according to the overhead minimization problem. The proposed DeepCQ+ routing technique with tuned reward function is referred to as DeepCQ+ routing policy. The DeepCQ+ routing algorithms are summarized in Algorithm \ref{DeepR2DN-algorithm}. 

\begin{algorithm}
\SetAlgoLined
\caption{The proposed DeepCQ+ routing}
\label{DeepR2DN-algorithm}
\small
{Receive incoming packet at node $i$:} \newline
\eIf{Packet is ACK}{Update $c$ and $h$ using (\ref{update_H}) and (\ref{update_C})}
    {\If{packet traversed a loop}{Drop packet, do not return ACK}
     \If{packet is already in queue}{
        Find best next-hop $i_1$ from (\ref{best_next_hop})) \\
        Compute $c_\text{ack}$ and $h_\text{ack}$ from (\ref{c_ack}) and (\ref{h_ack}) using $j^\star$ \\
        Drop packet but return ACK}
          \eIf{packet is not duplicate}
        {Add packet to the queue}
     {Do not add packet to the queue}}
    \If{ACK Packet is not received}{Do not update $c$ and $h$, (i.e. no packet is received)}
    \If{Queue is not empty}{Form the input to the DNN policy using (\ref{input_features}) and (\ref{preprocess_CH_sorting})
    DeepCQ+ routing: Choose $\begin{cases}
     \text{Broadcast} & \text{with probability } \pi_\theta(a = 1|\mathbf{o}_t; \theta) \\
     \text{Unicast} & \text{with probability } \pi_\theta(a = 0|\mathbf{o}_t; \theta)
     \end{cases}$}
\end{algorithm}

\section{Experiments and Numerical Results}
\subsection{Environment Modelling and Training Platforms}\label{sec:env_modelling}
We have developed an OpenAI gym \cite{brockman2016openai} environment in Python for training our DeepCQ+ agents. The python-based DeepCQ+ environment has all of the original CQ+ routing protocol features, including a plethora of MANET scenarios consisting of randomly generated dynamic network scenarios, multiple traffic flows, and mobility configurations. The environment is interfaced with the \textit{ray}, which is a powerful distributed DRL platform with RLlib library \cite{liang2018rllib} which provides scalable DRL training primitives with many built-in DRL algorithms such as PPO.

\begin{figure}[h]
\centering
\includegraphics[width=8.5cm]{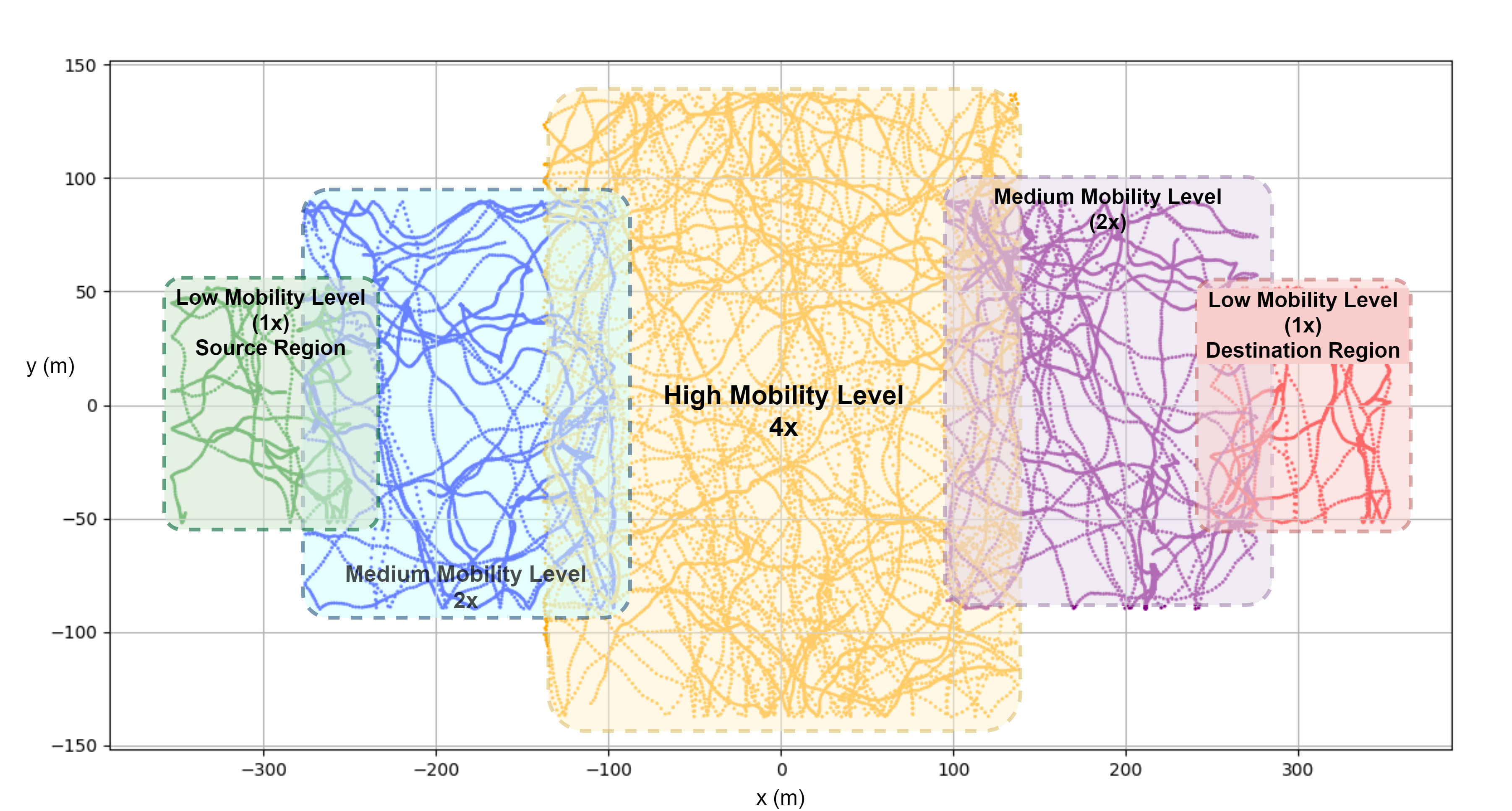}
\caption{Mobility regions and network topology. The movements of the nodes are shown for a network of size 30 according to the Gauss-Markov model}
\label{mobility}
\end{figure}

Within the MANET research community, there have been various mobility models proposed along with extensive discussions on their impact on the routing protocol performance \cite{gupta2013performance, Ariyakhajorn2006}. At a minimum, a MANET mobility model needs to have enough degree of randomization to cover corner cases and extreme scenarios required for extensive training in MADRL. The random way-point mobility model is often used in the simulation study of MANET, despite some unrealistic movement behaviors, such as exhibiting sudden stops and sharp turns \cite{Ariyakhajorn2006}. The Gauss-Markov mobility model, which has more realistic movement behavior and avoids often sudden stops and sharp returns, has less sensitivity of the network performance metrics with respect to the various randomness \cite{Ariyakhajorn2006}. 
To study the performance of MADRL under realistic mobility scenarios, our network scenario region is divided into 5 groups symmetrically between the source and destination nodes. The closer the region is to the center, the faster the nodes move. The central region has double the speed variance compard to variance of the mid-left and mid-right regions. The source and destination regions have half the speed variance compared to the mid-right/left regions. The mobility of the MANET nodes is simulated and shown in Fig. \ref{mobility}.
\begin{table}[h]
\begin{center}
\caption{List of configuration parameters and training hyperparameters}
 \begin{tabular}{r l l} 
 \hline
 \textbf{Parameters} & \textbf{Value in Training} & \textbf{Value in Test} \\ [0.5ex] 
 \hline\hline
 Network Sizes & 12 & 5 to 50\\ 
 \hline
 Learning Rate & 0.00005 & -\\
 \hline
 Discount Factor $\gamma$ & 0.99 & -\\
 \hline
 Episode Length & 3000 & 3000 \\
 \hline
 Region Size Scale & 1 & 2 \\
 \hline
Dynamic Level Scale & 1 & 5\\
\hline 
 Policy NN & FCNN(16, 8, 8, 4) & FCNN(16, 8, 8, 4)\\
 \hline
 Number of Data Flows & 1  & 1 to 4 \\
 \hline \hline
\end{tabular}
\label{tab:hyperparameters}
\end{center}
\end{table}
\subsection{Numerical Results}
\begin{figure*}[h]
\centering
\includegraphics[width=\textwidth]{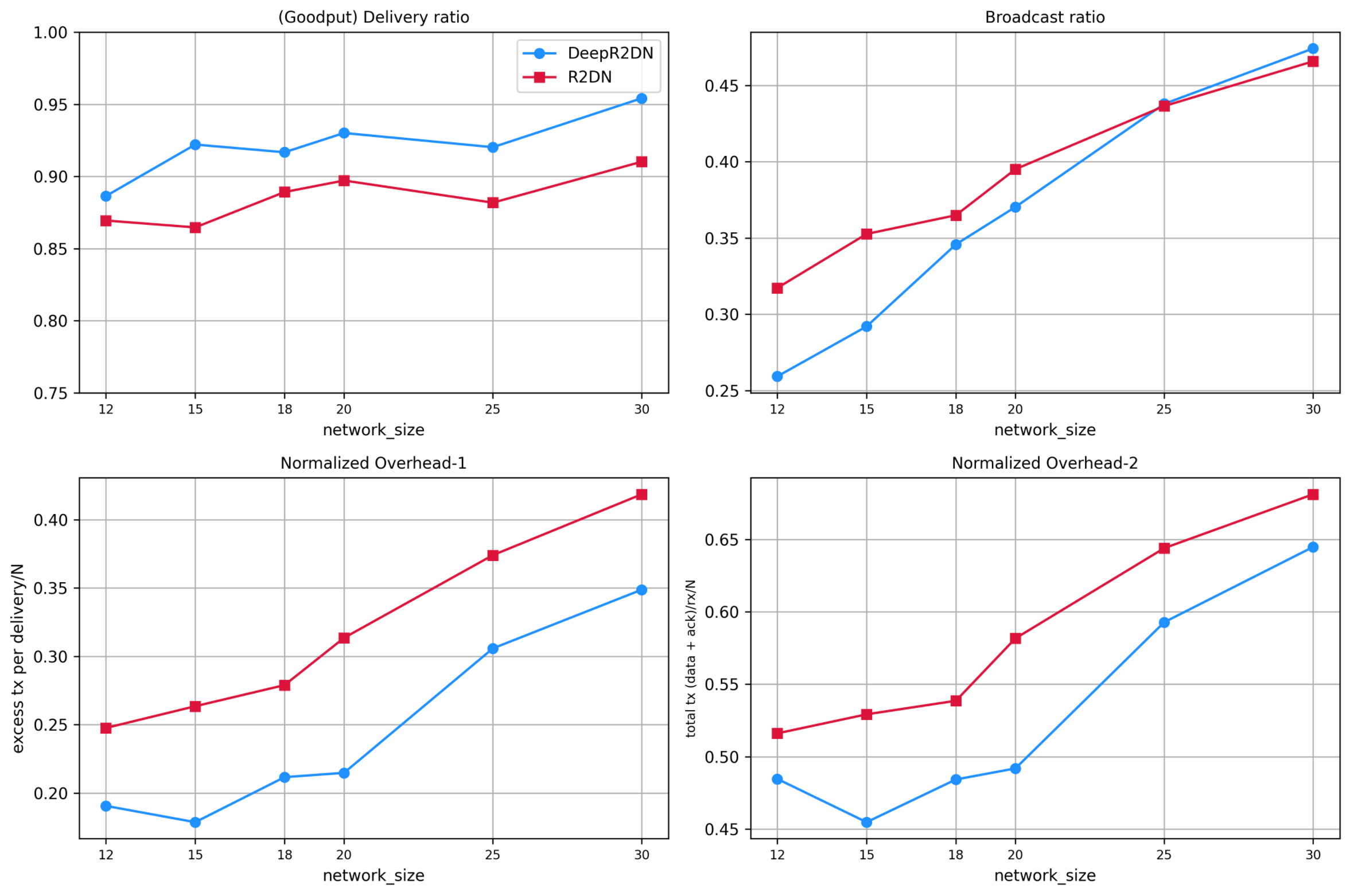}
\caption{Comparison of the results of the DeepCQ+ routing trained for a 12-node network only versus CQ+ routing; The results are tested across various network sizes from 12 to 30. Although the DeepCQ+ routing PPO policy is trained on 12-node networks, it scales perfectly for various network sizes. The DeepCQ+ routing achieves significantly lower normalized (divided by network size) overhead types 1 and 2 and higher delivery ratio and lower broadcast rates. }
\label{fig:deepVSCQ+}
\end{figure*}

Training and testing range of network parameters are summarized in Table \ref{tab:hyperparameters} with the hyperparameters used. Our training is performed over 50 million steps (approximately 15,000 episodes) and tests (where the results are shown in Fig. \ref{fig:deepVSCQ+}) are performed over network sizes between 10 and 30 with larger dynamic level and region scale randomization compared to the training. Training is performed on single network size (e.g. $N = 12$) scenarios. In our simulation experiments, $N = 12$ was the smallest size network in which the training results was extended to $N = 30$.  Fig. \ref{fig:deepVSCQ+} also shows the scalability and robustness of our training framework as the performance of the designed routing policies are maintained even with larger network parameters and configurations and networks of as large as 30 (which are not trained for). As shown in the figure, DeepCQ+ routing policy improves the goodput rate (delivery ratio) almost the same as the CQ+ routing but with significantly less overhead (both type 1 and type 2) for DeepCQ+. Indeed, our MADRL framework can adapt to different deployment tuples of goodput rate, broadcast rate, and even delay, which was not available in the CQ+ routing. Our results show that we require 10-25\% less overhead while achieves higher (1-5\%) delivery ratio (goodput rate), which shows significant improvement over the baseline CQ+ routing algorithm. DeepCQ+ also achieves lower percentage of broadcast transmissions, leading to network resource saving.  This overhead saving is attributed only to the policy's efficient decision on broadcast actions, while the next-hop is still selected based on the same CQ+ routing algorithm as in [\ref{best_next_hop}]. 

\section{Conclusions and Future Directions}
In this paper, we have presented and successfully demonstrated a novel MADRL-based CQ+ routing protocol which yields a robust, reliable, efficient, and scalable policy for dynamic wireless communication networks, including for scenarios that the algorithm was \emph{not} trained for. 
It is shown that DeepCQ+ is much more efficient than CQ+ routing techniques, significantly decreasing the required overhead for unit delivery. Moreover, the policy is scalable and uses parameter sharing for all nodes obtained during the training, which allows it to reuse the same trained policy for scenarios with different network configurations. It is noted that the sharing of parameters for all the nodes is not required during execution.

We are currently expanding the action space of DeepCQ+ to include next-hop selection for the unicast mode and other emerging MADRL techniques. Note that the focus of this work was on broadcast or unicast action selection while the next-hop was chosen based on the CQ+ routing. We plan to use recurrent neural network units as policy which is expected to capture higher gains compared to our current approach. 
\section{Acknowledgement}
Research reported in this publication was supported in part by Office of the Naval Research under the contract
N00014-19-C-1037. The content is solely the responsibility of the authors and does not necessarily represent the official views of the Office of Naval Research. The authors would like to thank Dr. Santanu Das (ONR Program Manager) for his support and encouragement. Also, we would like to thank the reviewers for their valuable comments to improve the quality of this paper.

\medskip

\small
\bibliographystyle{IEEEtran}
\bibliography{DeepR2DN.bib}

\end{document}